\documentclass[11pt,a4paper]{article}
\usepackage{amsmath,amssymb}

\usepackage{amsmath}

\usepackage{color}
\usepackage[colorlinks,linkcolor=blue,citecolor=red]{hyperref}

\DeclareMathSymbol{\bbbr}{\mathalpha}{AMSb}{"52}
\DeclareMathSymbol{\bbbc}{\mathalpha}{AMSb}{"52}

\newcommand{\R}{\mathbb{R}}

\begin{document}

\title{Quasilinear systems of Jordan block type and the mKP hierarchy}

\author{Lingling Xue$^{1}$, E.V. Ferapontov$^{2, 3}$
}
     \date{}
     \maketitle
     \vspace{-5mm}
\begin{center}
$^1$ Department of Mathematics\\ Ningbo University\\ Ningbo 315211, P.R. China \\
$^2$Department of Mathematical Sciences \\ Loughborough University \\
Loughborough, Leicestershire LE11 3TU \\ United Kingdom \\
$^3$Institute of Mathematics, Ufa Federal Research Centre,\\
Russian Academy of Sciences, 112, Chernyshevsky Street, \\ Ufa 450077, Russia \vspace{5pt}\\
\ \\
e-mails: \\[1ex]
\texttt{xuelingling@nbu.edu.cn}\\
\texttt{E.V.Ferapontov@lboro.ac.uk}\\
\end{center}

\vspace{1cm}

\begin{abstract}
We demonstrate that commuting quasilinear systems of Jordan block type are parametrised by solutions of the modified KP hierarchy. Systems of this form naturally occur as hydrodynamic reductions of multi-dimensional linearly degenerate dispersionless integrable PDEs.

\bigskip


\bigskip

\end{abstract}

\newpage



\tableofcontents

\section{Introduction}

Quasilinear systems of the form
\begin{equation}\label{q}
u_t=v(u)u_x
\end{equation}
have been thoroughly investigated in the literature. Here $u=(u^1, \dots, u^n)^T$ is a column vector of the dependent variables and $v$ is a $n\times n$ matrix.  The main emphasis has always been on the strictly hyperbolic  case where the matrix $v$ has real distinct eigenvalues. Under the additional condition that the Haantjes tensor of matrix $v$ vanishes, any such system can be reduced to the diagonal  form
\begin{equation}
R^i_t=\lambda^i(R) R^i_x,
\label{R}
\end{equation}
$i=1, \dots, n$, in specially adapted coordinates $R^1, \dots, R^n$ known as Riemann invariants. Systems of  type (\ref{R}) govern a wide range of problems in pure and applied mathematics, see e.g. \cite{Tsarev, Dub, Serre}.   It was shown by Tsarev \cite{Tsarev} that under the so-called semi-Hamiltonian constraint,
$$
 \left(\frac{\lambda
^i_j}{\lambda^j-\lambda^i}\right)_k=\left(\frac{\lambda
^i_k}{\lambda^k-\lambda^i}\right)_j,
$$
system (\ref{R}) possesses infinitely many conservation laws and commuting flows, and can be solved by the generalised hodograph method (here $i\ne j\ne k$ and low indices indicate differentiation by the variables $R^j$).

In this paper we study quasilinear systems  (\ref{q}) of Jordan block type. More precisely, we assume the existence of special coordinates (which we will also denote $R^1, \dots, R^n$) where the equations reduce to  upper-triangular Toeplitz form
\begin{equation}\label{J}
R_t=(\lambda^0E+\sum_{i=1}^{n-1}\lambda^iP^i)R_x;
\end{equation}
here $R=(R^1, \dots, R^n)^T$,  $E$ is the $n\times n$ identity matrix, $P$ is the $n\times n$ Jordan block with zero eigenvalue (note that $P^n=0$), and $\lambda^0, \lambda^i$ are  functions of $R$. Explicitly, a three-component version of system (\ref{J}) is as follows:
$$
\left(\begin{array}{c}R^1\\ R^2\\R^3\end{array}\right)_t=
\left(\begin{array}{ccc}
\lambda^0 & \lambda^1 & \lambda^2\\
0&\lambda^0&\lambda^1\\
0&0&\lambda^0
\end{array}\right)
\left(\begin{array}{c}R^1\\ R^2\\ R^3\end{array}\right)_x.
$$
The main properties of systems  (\ref{J}) can be summarised as follows:

\noindent (a) the corresponding matrix $v$ is pointwise of Jordan block type;

\noindent (b) the Haantjes tensor of matrix $v$ vanishes.

The vanishing of the Haantjes tensor makes  systems (\ref{J}) natural parabolic analogues of hydrodynamic type systems (\ref{R}) in Riemann invariants. Note that upper-triangular Toeplitz matrices  form a commutative family (cyclic Haantjes algebra in the terminology of \cite{TT}).
Systems of type (\ref{J}) appear as degenerations of hydrodynamic type systems associated with multi-dimensional hypergeometric functions  \cite{KK}, in the context of parabolic regularisation of the Riemann equation \cite{KO2}, and as   reductions of hydrodynamic chains and linearly degenerate dispersionless PDEs in 3D \cite{Pavlov1}. The most well-studied case of system (\ref{J}) corresponds to the choice $\lambda^0=R^1, \ \lambda^1=1, \ \lambda^i=0,\ i\geq 2$.

In section \ref{sec:J} we classify commuting systems of type (\ref{J}). Our main observation is that integrable hierarchies of Jordan block type are governed by the modified Kadomtsev-Petviashvili (mKP) hierarchy. Here is a brief summary of our results in this direction. Any two-component hierarchy of type (\ref{J}) can be parametrised in the form
$$
\left(\begin{array}{c}R^1\\ R^2\end{array}\right)_t=
\left(\begin{array}{cc}
\psi & \psi_1 \\
0&\psi
\end{array}\right)
\left(\begin{array}{c}R^1\\ R^2\end{array}\right)_x
$$
where $\psi$ satisfies the Lax equation of the mKP hierarchy,
$$
\psi_2=\psi_{11}+\rho\psi_1;
$$
here low indices indicate differentiation by $R^1, R^2$. Fixing the potential $\rho$ and varying  $\psi$ we obtain  commuting flows of the corresponding hierarchy. Similarly,
any three-component hierarchy of type (\ref{J}) can be parametrised in the form
$$
\left(\begin{array}{c}R^1\\ R^2\\R^3\end{array}\right)_t=
\left(\begin{array}{ccc}
\psi & \psi_1 & \psi_{11}+w_1\psi_1\\
0&\psi&\psi_1\\
0&0&\psi
\end{array}\right)
\left(\begin{array}{c}R^1\\ R^2\\ R^3\end{array}\right)_x,
$$
where $w$ solves the mKP equation
$$
4 w_{13}+6 w_1^2 w_{11}-w_{1111}-3w_{22}-6 w_{2}w_{11}=0,
$$
 and $\psi$ satisfies the corresponding Lax equations
\begin{equation*}
\psi_{2}={\psi_{11}}+2w_1\psi_{1},
\qquad 
\psi_{3}={\psi_{111}}+{3w_1}\psi_{11}
+\frac{3}{2}(w_{2}+w_{11}+w_1^2)\psi_{1}.
\end{equation*}
Fixing $w$ and varying $\psi$ we obtain commuting flows of the hierarchy. We show that the corresponding conserved densities are governed by the adjoint Lax equations.

In section \ref{sec:red} we demonstrate that
systems of Jordan block type naturally occur as hydrodynamic reductions of multi-dimensional linearly degenerate PDEs: the 3D Mikhalev system \cite{Mikhalev} is used as an illustrating example, see also \cite{Pavlov1}.

\section{Quasilinear systems of Jordan block type}
\label{sec:J}

\subsection{Form-invariance}

The class of two-component Toeplitz systems (\ref{J}),
$$
\left(\begin{array}{c}R^1\\ R^2\end{array}\right)_t=
\left(\begin{array}{cc}
\lambda^0 & \lambda^1 \\
0&\lambda^0\\
\end{array}\right)
\left(\begin{array}{c}R^1\\ R^2\end{array}\right)_x,
$$
is form-invariant under triangular changes of variables $(R^1, R^2)\leftrightarrow (r^1, r^2)$ of the form
\begin{equation}\label{sym2}
 R^1=F(r^1, r^2), \qquad  R^2=G(r^2),
\end{equation}
where $F$ and $G$ are arbitrary functions of the indicated arguments. Similarly, the class of three-component Toeplitz systems (\ref{J}),
$$
\left(\begin{array}{c}R^1\\ R^2\\R^3\end{array}\right)_t=
\left(\begin{array}{ccc}
\lambda^0 & \lambda^1 & \lambda^2\\
0&\lambda^0&\lambda^1\\
0&0&\lambda^0
\end{array}\right)
\left(\begin{array}{c}R^1\\ R^2\\ R^3\end{array}\right)_x,
$$
is form-invariant under triangular changes of variables $(R^1, R^2, R^3)\leftrightarrow (r^1, r^2, r^3)$ of the following form:
\begin{equation}\label{sym3}
R^1=r^1\frac{(\partial_{r^2}G)^2}{\partial_{r^3}H}+F(r^2, r^3), \qquad R^2=G(r^2, r^3), \qquad R^3=H(r^3),
\end{equation}
where $F, G$ and $H$ are arbitrary functions of the indicated arguments. This transformation freedom will be utilised to simplify the classification results. Note that the group preserving the class of diagonal systems (\ref{R}) is far more narrow, generated by transformations of the form $R^i=F^i(r^i)$,  functions of one variable only.

\subsection{Commuting flows}

Let us recall that commuting flows of  the diagonal system (\ref{R}) are governed by the equations
\begin{equation}\label{comm}
w^i_j=a^{ij}(w^j-w^i)
\end{equation}
where $a^{ij}(R)={\lambda^i_j}/{(\lambda^j-\lambda^i)}$ are fixed and solutions $w^i$ to linear system (\ref{comm}) vary \cite{Tsarev}. Note that the consistency conditions of system (\ref{comm}),
$$
a^{ij}_k=a^{ij}a^{jk}+a^{ik}a^{kj}-a^{ik}a^{ij},
$$
are equivalent to the integrable $2+1$-dimensional $n$-wave system. Thus, commuting flows of diagonal form (\ref{R}) are governed by the $n$-wave hierarchy.

Below we demonstrate that commuting flows of Toeplitz type (\ref{J}) are governed by the mKP hierarchy. Furthermore, the corresponding conserved densities of hydrodynamic type satisfy the adjoint Lax equations.

\subsubsection{Two-component case}

Direct calculation shows that the compatibility (commutativity) of two-component systems of the form
\begin{equation*}\label{2x2}
\left(\begin{array}{c}R^1\\ R^2\end{array}\right)_t=
\left(\begin{array}{cc}
\lambda^0 & \lambda^1 \\
0&\lambda^0
\end{array}\right)
\left(\begin{array}{c}R^1\\ R^2\end{array}\right)_x, \qquad
\left(\begin{array}{c}R^1\\ R^2\end{array}\right)_y=
\left(\begin{array}{cc}
\mu^0 & \mu^1 \\
0&\mu^0
\end{array}\right)
\left(\begin{array}{c}R^1\\ R^2\end{array}\right)_x
\end{equation*}
is equivalent to the following two conditions:
$$
\frac{\lambda^0_1}{\lambda^1}=\frac{\mu^0_1}{\mu^1},
$$
$$
 \frac{\lambda^0_2-\lambda^1_1}{\lambda^1}=\frac{\mu^0_2-\mu^1_1}{\mu^1}.
$$
Let us introduce the notation $\frac{\lambda^0_1}{\lambda^1}=m, \
 \frac{\lambda^0_2-\lambda^1_1}{\lambda^1}=\rho$; note that the quantities $m, \rho$ are shared by all commuting flows.
 If $m\ne 0$, using  symmetry (\ref{sym2}) one can set $m=1$. Denoting $\lambda^0=\psi$ we obtain
 $\lambda^1=\psi_1, \ \psi_2=\psi_{11}+\rho\psi_1$. Thus, members of the commuting hierarchy can be parametrised in the form
\begin{equation}\label{psi}
\left(\begin{array}{c}R^1\\ R^2\end{array}\right)_t=
\left(\begin{array}{cc}
\psi & \psi_1 \\
0&\psi
\end{array}\right)
\left(\begin{array}{c}R^1\\ R^2\end{array}\right)_x,
\end{equation}
where $\psi$ solves the Lax equation of the mKP hierarchy,
$$
\psi_2=\psi_{11}+\rho\psi_1.
$$
Here the `potential' $\rho$ is fixed and the solution $\psi$ of the Lax equation varies.
The choice $\rho=0,\ \psi=R^1$ gives rise to the system
\begin{equation*}
\left(\begin{array}{c}R^1\\ R^2\end{array}\right)_t=
\left(\begin{array}{cc}
R^1 & 1 \\
0&R^1
\end{array}\right)
\left(\begin{array}{c}R^1\\ R^2\end{array}\right)_x
\end{equation*}
whose commuting flows are parametrised by solutions to the heat equation \cite{KK, Pavlov1}.

\medskip

\noindent {\bf Remark 1.} Conservation laws of system (\ref{psi}) are relations of the form
$\phi_t=g_x$ which hold identically modulo (\ref{psi}). This gives
$$
g_1=\psi \phi_1, \quad g_2=\psi \phi_2+\psi_1 \phi_1,
$$
and the elimination of $g$ results in the adjoint Lax equation for the conserved density $\phi$:
$$
\phi_2=-\phi_{11}+\rho \phi_1.
$$

\subsubsection{Three-component case}

Direct calculation shows that the compatibility  of three-component systems of the form
$$
\left(\begin{array}{c}R^1\\ R^2\\R^3\end{array}\right)_t=
\left(\begin{array}{ccc}
\lambda^0 & \lambda^1 & \lambda^2\\
0&\lambda^0&\lambda^1\\
0&0&\lambda^0
\end{array}\right)
\left(\begin{array}{c}R^1\\ R^2\\ R^3\end{array}\right)_x, \qquad
\left(\begin{array}{c}R^1\\ R^2\\R^3\end{array}\right)_y=
\left(\begin{array}{ccc}
\mu^0 & \mu^1 & \mu^2\\
0&\mu^0&\mu^1\\
0&0&\mu^0
\end{array}\right)
\left(\begin{array}{c}R^1\\ R^2\\ R^3\end{array}\right)_x
$$
is equivalent to the following six conditions (we assume $\lambda^1\neq 0$ and $\mu^1\neq 0$):
\begin{equation}\label{eq_m}
m\equiv\frac{\lambda^0_{1}}{\lambda^{1}} =\frac{\mu^{0}_{1}}{\mu^{1}},
\end{equation}
\begin{equation}\label{eq_q}
q\equiv\frac{\lambda^{1}_{1}-m\lambda^{2}}{\lambda^{1}} =\frac{\mu^{1}_{1}-m\mu^{2}}{\mu^{1}},
\end{equation}
\begin{equation}\label{eq_p}
p\equiv\frac{\lambda^0_{2}-m\lambda^{2}}{\lambda^{1}} =\frac{\mu^{0}_{2}-m\mu^{2}}{\mu^{1}},
\end{equation}
\begin{equation}\label{eq_s}
s\equiv\frac{\lambda^{1}_{2}-\lambda^{2}_{1}-p \lambda^{2}}{\lambda^{1}} =\frac{\mu^{1}_{2}-\mu^{2}_{1}-p \mu^{2}}{\mu^{1}},
\end{equation}
\begin{equation}\label{eq_r}
r\equiv\frac{\lambda^0_{3}-\lambda^{1}_{2}+q \lambda^{2}}{\lambda^{1}} =\frac{\mu^{0}_{3}-\mu^{1}_{2}+q \mu^{2}}{\mu^{1}},
\end{equation}
\begin{equation}\label{eq_h}
h\equiv\frac{\lambda^{1}_{3}-\lambda^{2}_{2}-(s+r) \lambda^{2}}{\lambda^{1}} =\frac{\mu^{1}_{3}-\mu^{2}_{2}-(s+r) \mu^{2}}{\mu^{1}}.
\end{equation}
These conditions can be reduced to the mKP equation and its Lax pair as follows. Assuming $m\neq0$ and denoting $\lambda^0=\psi$, from \eqref{eq_m} and \eqref{eq_q} one obtains
\begin{equation*}
\lambda^{1}=\frac{\psi_{1}}{m},
\quad
\lambda^{2}=\frac{\psi_{11}}{m^2}-\left(\frac{q}{m^2}+\frac{m_1}{m^3}\right) \psi_{1}.
\end{equation*}
Inserting the above into \eqref{eq_p} yields
\begin{equation*}
\psi_{2}=\frac{\psi_{11}}{m}+\left(\frac{p-q}{m}-\frac{m_1}{m^2}\right) \psi_{1},
\end{equation*}
so that \eqref{eq_s} gives
\begin{equation*}
\frac{m_1}{m}\psi_{11}-\left[sm-pq-p_1+m_2+\frac{m_1}{m}q+\left(\frac{m_1}{m}\right)^2\right]\psi_1=0.
\end{equation*}
This implies
\[m_1=0, \quad s=\frac{1}{m}(q p+p_{1}-m_2).\]
As $m\ne 0$, modulo transformations (\ref{sym3}) one can set
$m=1,\ s=q p+p_{1},$
so that the formulae for $\lambda^0, \lambda^1, \lambda^2$ simplify to
$\lambda^0=\psi, \ \lambda^1=\psi_1, \ \lambda^2=\psi_{11}-q\psi_1$
where
\begin{equation}\label{eq_p1}
 \qquad \psi_2=\psi_{11}+(p-q)\psi_1.
\end{equation}
Then \eqref{eq_r} yields
\begin{equation}\label{eq_r1}
\psi_{3}={\psi_{111}}+{(p-2q)}\psi_{11}
+\big[q^2+r+(p-q)_1\big]\psi_{1}.
\end{equation}
Finally, it follows from \eqref{eq_h} that
\begin{equation*}
(p+q)_{1}{ \psi_{11}}-\left(q^2p-h+rq+ q_{2}+q q_{1} +2 q p_{1}+ r_{1}\right) \psi_{1}=0,
\end{equation*}
which allows us to set
\begin{equation*}
q=-p,\quad
h=(p)^3-rp- p_{2}-p p_{1}+ r_{1}.
\end{equation*}
The compatibility of  \eqref{eq_p1} and \eqref{eq_r1}, i.e., the condition $\psi_{23}=\psi_{32}$,
leads to
\begin{equation*}
\big[p_{11}-2pp_{1}+2r_{1}-3p_{2}\big]\psi_{11}
=\big[
2p_{1}^2+2(r-p^2)p_{1}-2p r_{1}+2pp_{2}-r_{11}+r_{2}+2 p_{12}-2p_{3}
\big]\psi_{1},
\end{equation*}
from which we obtain the following two equations:
\begin{equation}\label{eq_p11}
p_{11}-2pp_{1}+2r_{1}-3p_{2}=0,
\end{equation}
\begin{equation}\label{eq_mkp0}
2p_{1}^2+2(r-p^2)p_{1}-2p r_{1}+2pp_{2}-r_{11}+r_{2}+2 p_{12}-2p_{3}=0.
\end{equation}
In order to solve  \eqref{eq_p11} for $r$ we introduce the potential variable $w$ such that $p= w_{1}$. Then integrating \eqref{eq_p11} gives
\[r=\frac{1}{2}w_1^2-\frac{1}{2} w_{11}+\frac{3}{2}w_{2}.\]
Finally from \eqref{eq_mkp0} we obtain
\begin{equation}\label{eq_mkp}
4 w_{13}+6 w_1^2 w_{11}-w_{1111}-3w_{22}-6 w_{2}w_{11}=0,
\end{equation}
which is the potential  mKP equation.
Note that the mKP hierarchy was  introduced in \cite{K1},  \cite{JM}. The generalised Miura transformation connecting KP and mKP equations was constructed in \cite{K1}, see also \cite{K2} for the first classification results of integrable equations in 2+1 dimensions. Exact solutions of mKP equation were constructed in \cite{K3, G1, G2}, see also references therein.

To summarise, members of the three-component commuting hierarchy can be parametrised in the form
\begin{equation}\label{psi3}
\left(\begin{array}{c}R^1\\ R^2\\R^3\end{array}\right)_t=
\left(\begin{array}{ccc}
\psi & \psi_1 & \psi_{11}+w_1\psi_1\\
0&\psi&\psi_1\\
0&0&\psi
\end{array}\right)
\left(\begin{array}{c}R^1\\ R^2\\ R^3\end{array}\right)_x
\end{equation}
where $w$ satisfies the mKP equation (\ref{eq_mkp}) and $\psi$ solves the corresponding Lax equations  \eqref{eq_p1}, \eqref{eq_r1}:
\begin{equation}\label{mkp_lax}
\psi_{2}={\psi_{11}}+2w_1\psi_{1},
\quad 
\psi_{3}={\psi_{111}}+{3w_1}\psi_{11}
+\frac{3}{2}(w_{2}+w_{11}+w_1^2)\psi_{1}.
\end{equation}
Fixing $w$ and varying $\psi$ we obtain commuting flows of the hierarchy.

\medskip

\noindent{\bf Example.} Set $w=0$, then equations for $\psi$ and $\lambda^0, \ \lambda^1, \ \lambda^2$ become
\begin{equation*}
\psi_{2}={\psi_{11}},
\quad 
\psi_{3}={\psi_{111}},\qquad
\lambda^0=\psi, \quad \lambda^{1}={\psi_{1}},
\quad
\lambda^{2}={\psi_{11}}.
\end{equation*}
We can choose
$$
\psi=R^{1},\quad \lambda^0=R^1, \quad \lambda^{1}=1,\quad \lambda^{2}=0,
$$
or
$$
\psi=e^{k R^{1}+k^2 R^{2}+k^3 R^{3}},\quad \lambda^0=\psi, \quad \lambda^{1}=k \psi,\quad \lambda^{2}=k^2 \psi,
$$
where $k$ is an arbitrary constant
(the former solution was  considered in \cite{Pavlov1}).

\medskip

\noindent {\bf Remark 2.} Conservation laws of  system  (\ref{psi3}) are relations of the form
$\phi_t=g_x$  which hold identically modulo (\ref{psi3}). This gives
$$
g_1=\psi \phi_1, \quad g_2=\psi \phi_2+\psi_1 \phi_1,\quad g_3=\psi \phi_3+\psi_1 \phi_2+(\psi_{11}+w_1\psi_1)\phi_{1}.
$$
The elimination of $g$ results in the adjoint Lax equations for the conserved density $\phi$:
\begin{equation*}
\phi_{2}=-{\phi_{11}}+2w_1\phi_{1},
\quad 
\phi_{3}={\phi_{111}}-{3w_1}\phi_{11}
+\frac{3}{2}(w_{2}-w_{11}+w_1^2)\phi_{1}.
\end{equation*}




\subsubsection{Four-component case}

Omitting details of calculations we present the final result: four-component commuting flows of  Toeplitz type (\ref{J})   can be  parametrised in the form
\begin{equation}\label{4}
{\small \left(\begin{array}{c}R^1\\ R^2\\R^3\\R^4\end{array}\right)_t=
\left(\begin{array}{cccc}
\psi & \psi_1 & \psi_{11}+w_1\psi_1 & \psi_{111}+ 2w_{1}\psi_{11}+ \frac{1}{2}(w_{2}+3w_{11}+w_{1}^2)\psi_1
\\
0&\psi&\psi_1&\psi_{11}+w_1\psi_1\\
0&0&\psi&\psi_1\\
0&0&0&\psi
\end{array}\right)
\left(\begin{array}{c}R^1\\ R^2\\ R^3\\R^4\end{array}\right)_x}
\end{equation}
where $w$ solves the first three equations of the mKP hierarchy,
\begin{equation*}
4 w_{13}+6 w_1^2 w_{11}-w_{1111}-3w_{22}-6 w_{2}w_{11}=0,
\end{equation*}
$$
2  w_{23} = 3 w_{14}-w_{1112}+(3 w_{1}^2-3 w_{2}) w_{12}+6 w_{1} w_{11} w_{2}+(w_{111}-4 w_{3}-2 w_{1}^3) w_{11},
$$
$$
w_{24} = \frac{8}{9} w_{33}-\frac{1}{9} w_{111111}+\frac{2}{9} w_{1113}
-2 w_{1} w_{2} w_{12}
+4 w_{111} w_{1} w_{11}+(w_{1}^2-w_{2}) w_{1111}
$$
$$\quad\quad
+\frac{2}{3}( w_{1}^3+2 w_{3}-2 w_{111}) w_{12}
+\frac{4}{3} w_{11}^3+(4 w_{1}^2 w_{2}-2 w_{1}^4-2 w_{2}^2-w_{112}) w_{11},
$$
and $\psi$ satisfies the corresponding Lax equations:
\begin{equation*}
\begin{array}{c}
\psi_{2}={\psi_{11}}+2w_1\psi_{1},
\quad 
\psi_{3}={\psi_{111}}+{3w_1}\psi_{11}
+\frac{3}{2}(w_{2}+w_{11}+w_1^2)\psi_{1},\\
\ \\
\psi_{4}={\psi_{1111}}+{4w_1}\psi_{111}
+(2w_2+4w_{11}+4w_1^2)\psi_{11}
+\Delta \psi_{1};
\end{array}
\end{equation*}
here
$$
\Delta=\frac{4}{3} w_{3}+\frac{5}{3}w_{111}+\frac{2}{3}w_{1}^3+4w_{1}w_{11}+w_{12}+2 w_{1} w_{2}.
$$
Fixing $w$ and varying $\psi$ we obtain commuting flows of the hierarchy.

\medskip

\noindent {\bf Remark 3.} For the matrix elements of (\ref{4}),
$$
\lambda^0=\psi, \quad \lambda^1=\psi_1, \quad \lambda^2=\psi_{11}+w_1\psi_1, \quad \lambda^3=\psi_{111}+ 2w_{1}\psi_{11}+ \frac{1}{2}(w_{1}^2+w_{2}+3w_{11})\psi_1,
$$
we have simple recursive formulae:
$$
\lambda^1=\psi_1, \quad \lambda^2+w_1\lambda^1=\psi_2, \quad \lambda^3+w_1\lambda^2+w_2\lambda^1=\psi_3.
$$
This recurrence generalises to the general $n$-component case:
$$
\lambda^k+w_1\lambda^{k-1}+w_2\lambda^{k-2}+\dots+w_{k-1}\lambda^1=\psi_k, \quad 1\leq k\leq n-1.
$$


\subsection{Generalised hodograph method}

Solutions to system (\ref{J}) can be obtained by the following recipe which is analogous to the generalised hodograph method of Tsarev \cite{Tsarev}. Let
\begin{equation*}
R_y=(\mu^0E+\sum_{i=1}^{n-1}\mu^iP^i)R_x
\end{equation*}
be a commuting flow of system (\ref{J}). Then the matrix equation
$$
\mu^0E+\sum_{i=1}^{n-1}\mu^iP^i=Ex+(\lambda^0E+\sum_{i=1}^{n-1}\lambda^iP^i)t
$$defines an implicit solution of (\ref{J}). In components, this is equivalent to $n$ implicit relations
$$
\mu^0=x+\lambda^0t, \quad \mu^1=\lambda^1t, \quad \dots,\quad  \mu^n=\lambda^n t.
$$

\section{Jordan type reductions of linearly degenerate PDEs}
\label{sec:red}

It is remarkable that, although for `strongly nonlinear' PDEs such as the dispersionless KP/Toda equations, the Jordan type reductions do not occur,  they naturally arise in the context of multi-dimensional {\it linearly degenerate} PDEs (such as linearly degenerate systems of hydrodynamic type, Monge-Amp\'ere  equations, etc). Below we illustrate this phenomenon for the 3D Mikhalev system \cite{Mikhalev}.

\subsection{3D Mikhalev system}
\label{sec:M3}

Here we consider the system
\begin{equation}\label{Mik}
u_t=v_y+uv_x-vu_x, \qquad v_x=u_y.
\end{equation}

\noindent {\bf Two-component} hydrodynamic reductions (of Jordan block type) of system (\ref{Mik}) are exact solutions of the form
\begin{equation}\label{uw}
u=u(R^1, R^2), \quad v=v(R^1, R^2)
\end{equation}
where the variables $R^1, R^2$ satisfy a pair of commuting $2\times 2$ systems (\ref{2x2}):
\begin{equation*}\label{2x2}
\left(\begin{array}{c}R^1\\ R^2\end{array}\right)_t=
\left(\begin{array}{cc}
\psi & \psi_{1} \\
0&\psi
\end{array}\right)
\left(\begin{array}{c}R^1\\ R^2\end{array}\right)_x, \qquad
\left(\begin{array}{c}R^1\\ R^2\end{array}\right)_y=
\left(\begin{array}{cc}
\varphi & \varphi_{1} \\
0&\varphi
\end{array}\right)
\left(\begin{array}{c}R^1\\ R^2\end{array}\right)_x,
\end{equation*}
with
$$
\psi_2=\psi_{11}+\rho\psi_1,\quad \varphi_2=\varphi_{11}+\rho\varphi_1.
$$
All such reductions can be described explicitly. Direct calculation shows that $u$ and $v$ must be polynomial in $R^1$ of degree 2 and 4, respectively:
$$
u=-\frac{1}{4}a' (R^{1})^2+ bR^1+c, \quad v=-\frac{1}{2}u^2+au +d,
$$
where the coefficients $a, b, c, d$ are functions of $R^2$ satisfying a single relation $d'+b^2+ca'=0$ (prime denotes differentiation by $R^2$). The functions $\varphi, \psi$ and $\rho$ are expressed in terms of $u$ by the formulae
$$
\varphi=-u+a,\quad \psi=-v+a\varphi, \quad \rho =
\frac{2 u_{2}-a'}{2u_{1}}.
$$
A particular choice $a=0, \ b=-1,\  c=0, \  d=-R^2$ leads to
$$
u=-R^1, \quad v=-R^2-\frac{1}{2}(R^1)^2, \quad \varphi=R^1, \quad \psi=R^2+\frac{1}{2}(R^1)^2,\quad \rho=0,
$$
the case considered in \cite{Pavlov1}.

\medskip

\noindent {\bf Three component} reductions of Jordan block type can be sought in the form
\begin{equation*}
\displaystyle{
R_t=
\left(\begin{array}{ccc}
\psi & \psi_1 & \psi_{11}+{w}_1\psi_1\\
0&\psi&\psi_1\\
0&0&\psi
\end{array}\right)
R_{x},
\qquad
R_{y}=
\left(\begin{array}{ccc}
\varphi & \varphi_1 & \varphi_{11}+{w}_1\varphi_1\\
0&\varphi&\varphi_1\\
0&0&\varphi
\end{array}\right)
R_{x}
},
\end{equation*}
where ${w}$ satisfies the mKP equation (\ref{eq_mkp}) and  $\psi$, $\varphi$ are two solutions of the corresponding Lax equations  \eqref{mkp_lax}. In this case the formulae become more complicated. Direct calculation shows that $u$ and $v$ must be polynomial in $R^1, R^2$:
$$
u = b R^1-\frac{4}{27}a''(R^2)^3+\frac{2}{3}\gamma'(R^2)^2+\alpha R^2+\beta,
$$
$$
v = -\frac{1}{2} u^2+ au-\frac{4}{27} a'^2 (R^2)^3-b\gamma R^2+\zeta,\quad
$$
where $b=\gamma-\frac{2}{3} a' R^2$ and $a$, $\alpha$, $\beta$, $\gamma$, $\zeta$ are functions of $R^3$ with the condition
$
\zeta' +a'\beta+\frac{3}{2}\alpha\gamma=0.
$
The functions $\varphi, \psi$ and $w$ are as follows:
$$\varphi=-u+a,\quad \psi=-v+a\varphi ,
$$
$$
w = -\frac{a'}{ 6b}(R^1)^2
-\frac{1}{ 18b}R^1\left(4a''(R^2)^2-12\gamma'R^2
-9\alpha \right)+f,
$$
where $f$ is a  function of $R^2$ and $R^3$ which satisfies the equation
\begin{equation}\label{f}
 \frac{1}{2} (3 \gamma-2 a' R^2)^2 f_{2}=
x^4 (R^2)^4
+x^3 (R^2)^3+x^2 (R^2)^2+x^1 (R^2)+3x^0.
\end{equation}
Here the coefficients $x^i$ are functions of $R^3$ defined as
$$
x^4\equiv \frac{8}{ 27} a''' a'- \frac{2}{9} a''^2,\quad x^3\equiv
\frac{4}{ 3} a''\gamma'-\frac{4}{ 9} a''' \gamma-\frac{4}{ 3}\gamma'' a',
$$
$$
x^2\equiv \alpha a'' -2 \alpha' a'+2 \gamma'' \gamma-2\gamma'^2,\quad
x^1\equiv 3 \alpha' \gamma-2 \beta' a'-3 \alpha\gamma'+a'^2,\quad
x^0\equiv \beta' \gamma- \frac{1}{ 2} a' \gamma-\frac{3}{8} \alpha^2.
$$
If $a'=0$, then \eqref{f} gives 
$$
f =
\frac{4}{27}
   (R^2)^3 \left(\frac{\gamma'}{\gamma}   \right)'
+\frac{1}{3} (R^2)^2\left( \frac{\alpha  }{\gamma}  \right)'
+R^2\left(\frac{2\beta'}{3\gamma}  -\frac{\alpha ^2}{4\gamma ^2}  \right)  +\eta,
$$
otherwise,
$$
f= \frac{(R^2)^3}{6a'^2} x^{4}
+\frac{(R^2)^2}{ 4a'^3} (a' x^{3}+3 \gamma x^{4})
+\frac{R^2}{ 8 a'^4} (4 a'^2 x^{2}+12 a' \gamma x^{3}+27 \gamma^2 x^{4})
$$
$$
+\frac{1}{ 16 a'^5 b} (16 a'^4 x^{0}+8 a'^3 \gamma x^{1}+12 a'^2 \gamma^2 x^{2}
+18 a' \gamma^3 x^{3}+27 \gamma^4 x^{4})
$$
$$
+\frac{\ln{b}}{ 8 a'^5}  (4 a'^3 x^{1}+12 a'^2 \gamma x^{2}+27 a' \gamma^2 x^{3}
+54 \gamma^3 x^{4})+\eta;
$$
here $\eta$ is an extra arbitrary function of $R^3$.


\section*{Acknowledgements}

We thank Maxim Pavlov and Vladimir Novikov for useful comments. EVF also thanks Alexey Bolsinov and David Calderbank for a discussion on quasilinear systems of Jordan block type and their role  in the general method of hydrodynamic reductions. LX was supported by the National Natural Science Foundation of China (Grant No. 11501312) and the K.C. Wong Magna Fund in Ningbo University. LX also thanks Loughborough University for a kind hospitality.

\end{document}